\documentclass[preprint,prd,showpacs,superscriptaddress]{revtex4}
\usepackage{amssymb,amsmath,graphicx,amscd}
\usepackage[mathscr]{eucal}
\usepackage{verbatim,color,ulem}

\begin{document}

\title{ Quantum Modifications to Gravity Waves in de~Sitter Spacetime}
\author{Jen-Tsung Hsiang}\email{cosmology@gmail.com}
\affiliation{Department of Physics, National Dong Hwa University,
Hualien,
 Taiwan, ROC}
\author{L. H. Ford}\email{ford@cosmos,phy.tufts.edu}
\affiliation{Institute of Cosmology, Department of Physics and
Astronomy,
 Tufts University, Medford, MA 02155, USA}
\author{Da-Shin Lee}\email{dslee@mail.phys.ndhu.edu.tw}
\affiliation{Department of Physics, National Dong Hwa University,
Hualien,
Taiwan, ROC}
\author{Hoi-Lai Yu}\email{hlyu@phys.sinica.edu.tw}
\affiliation{Institute of Physics, Academia Sinica, Nankang,
 Taipei 11529, Taiwan, ROC}

\begin{abstract}
We treat a model in which tensor perturbations of de~Sitter spacetime,
represented as a spatially flat model,
are modified by the effects of the vacuum fluctuations of a massless
conformally invariant field, such as the electromagnetic field.
We use the semiclassical theory of gravity with the expectation value
of the conformal field stress tensor as a source. We first study the
stability  of de~Sitter spacetime by searching for growing, spatially
homogeneous modes, and conclude that it is stable within the limits of
validity of the semiclassical theory. We next examine the modification
of linearized plane gravity waves by the effects of the quantum
stress tensor. We find a correction term which is of the same form
as the original wave, but displaced in phase by $-\pi/2$, and with an
amplitude which depends upon the duration of inflation. The magnitude of
this effect is proportional to the change in scale factor during
inflation. So long as the energy scale of inflation and the proper
frequency of the mode at the beginning of inflation are well below
the Planck scale, the fractional correction is small. However, modes
which are transplanckian at the onset of inflation can undergo a
significant correction. The increase in amplitude can potentially
have observable consequences through a modification of the power
spectrum of tensor perturbations in inflationary cosmology. This
enhancement of the power spectrum depends upon the duration of
inflation and is greater for shorter wavelengths.
\end{abstract}


\maketitle

\baselineskip=13pt

\section{Introduction}

Most versions of inflationary cosmology assume a period of
exponential expansion in which the universe is approximately
a portion of de~Sitter spacetime. Quantum fields in  de~Sitter
spacetime play a crucial role in creating the primordial spectrum of
scalar and tensor perturbations. In addition, quantum effects can
potentially modify the duration of inflation and possibly introduce
instabilities. Recently, there has been work on the possible effects of
quantum stress tensor fluctuations in inflation~\cite{WKF07,FMNWW10}.

In the present paper, we examine some effects in the semiclassical
theory, where gravity is coupled to the renormalized expectation
value of a matter field stress tensor,
the mean value around which
stress tensor fluctuations occur. The semiclassical theory has been
extensively studied and applied to scalar perturbations of
de~Sitter spacetime. (See, for example, Ref.~\cite{AMM09}
and references therein.) There seems to have been less attention
paid to tensor perturbations, which will be the topic of this
paper. A brief discussion was given by Starobinsky~\cite{S81}
and a more detailed derivation of the equations for tensor
perturbations was given by Campos and Verdaguer~\cite{CV94}.
We will treat a model in which the matter field is a conformal
field, such as the electromagnetic field, and address two physical
questions: the stability of de~Sitter spacetime under tensor
perturbations, and the effects of one-loop quantum matter field
corrections upon the propagation of gravity waves in de~Sitter
spacetime.

In Sect.~\ref{sec:perturbed}, we review the aspects of the semiclassical theory
needed for our analysis. Section~\ref{sec:geom} treats the geometric terms
in the stress tensor expectation value. Here we find that these terms
have no physical effect for our problems. The stability of the
tensor perturbations is discussed in Sect.~\ref{sec:stab}. The one-loop
correction to gravity wave modes is derived in
Sect.~\ref{sec:correct}, and the possible implications for inflationary
cosmology are discussed in Sect.~\ref{sec:tensor-power}.
Our results are summarized in Sect.~\ref{sec:final}.

We adopt the sign conventions of Ref.~\cite{MTW}, and use units in which
$\hbar = c =1$.

\section{Weakly Perturbed De~Sitter Spacetime}
\label{sec:perturbed}

We will be concerned with the piece of global de~Sitter spacetime which
can be represented as a spatially flat Robertson-Walker universe with
the metric
\begin{equation}
ds^2 = a^2(\eta)\,(-d\eta^2 + dx^2 +dy^2 +dz^2)\,,  \label{eq:metric}
\end{equation}
where $a(\eta) = -1/(H \eta)$ and $\eta <0$ is the conformal time
coordinate. We wish to consider tensor perturbations of this geometry,
which describe gravitational waves on the de~Sitter background. Let
the perturbed metric be
\begin{equation}
g_{\mu\nu}= \gamma_{\mu\nu} + h_{\mu\nu} \,,
\end{equation}
where $\gamma_{\mu\nu}$ is the background metric of Eq.~(\ref{eq:metric}), and
$h_{\mu\nu}$ is the perturbation. We will employ the transverse
trace-free gauge defined by
\begin{equation}
h^{\mu\nu}_{~~~;\nu} = 0, ~~~h = 0 ~~~{\rm and~}~~~ h^{\mu\nu}u_\nu =
0\,.    \label{eq:TT}
\end{equation}
Here $u^\nu=\delta^\nu_t$ is the four velocity of the comoving observers,
covariant  derivatives are taken respect to the fixed de~Sitter
background, and indices are raised and lowered by the background metric.
These conditions remove all of the gauge freedom, and  leave only the
two physical degrees of freedom associated with the possible
polarizations of a gravity wave.

It was shown long ago by Lifshitz~\cite{LS}  that the mixed
components $ h_\mu
^\nu$ satisfy the {\it scalar} wave equation
\begin{equation}
\Box_s h_\mu ^\nu = 0\,, \label{eq:scalar}
\end{equation}
where $\Box_s$ is the scalar wave operator.
One consequence of this result is that
de~Sitter spacetime is classically stable against tensor perturbations,
as the solutions of Eq.~(\ref{eq:scalar}) are oscillatory functions.
A second consequence is that
gravitons in de~Sitter spacetime behave as a pair of massless,
minimally coupled quantum scalar fields~\cite{FP77}.

It is well known that such massless scalar fields exhibit a type of
quantum instability in that they do not possess a de~Sitter invariant
vacuum state. As a result, the mean squared field grows linearly
in time~\cite{VF82,Linde,Staro} as
$\langle \varphi^2 \rangle \sim {H^3 t}/(4 \pi^2)$.
Similarly, the mean squared graviton field also grows linearly:
$\langle  h_\mu ^\nu \,h_\nu ^\mu \rangle \sim {H^3 t}/{ \pi^2}$.
However, this growth does not produce any physical consequences, at
least in pure quantum gravity at the one loop level. It was shown
in Ref.~\cite{Ford} that this level, all of the linearly growing terms
cancel in the graviton effective energy momentum tensor. Whether
there is an instability at higher orders is still
unclear~\cite{TW96,GT08,TW08}.

In this paper we will study a model involving coupling of the tensor
perturbations to a matter field. As a prelude, let us briefly recall
the essential features of the renormalization of
$\langle  T_{\mu \nu} \rangle$, the expectation value
of a matter stress tensor on a curved background~\cite{BD82}.
This quantity is
formally divergent, but under a covariant regularization, the
divergent terms are of three types. The first is proportional to the
metric tensor, and can be absorbed in a cosmological constant
renormalization. The second is proportional to the Einstein tensor,
and can be absorbed in a renormalization of Newton's
constant. Finally, there are divergent terms proportional to two
geometric tensors, $H_{\mu\nu}$ and  $A_{\mu\nu}$, which arise from
variation of $R^2$ and $C_{\mu\nu\alpha\beta} C^{\mu\nu\alpha\beta}$
terms in the action, respectively. Here $R$ is the scalar curvature
and $C_{\mu\nu\alpha\beta}$ is the Weyl tensor. The explicit forms of
these tensors are expressible in terms of $R$, the Ricci tensor,
$R_{\mu\nu}$, and their second derivatives as
\begin{equation}
H_{\mu\nu} =  -2\nabla_\nu \nabla_\mu R + 2g_{\mu\nu}\nabla_\rho \nabla^\rho R
 - \frac{1}{2}g_{\mu\nu} R^2 +2R R_{\mu\nu}\,,  \label{eq:H}
\end{equation}
and~\cite{AMM03}
\begin{equation}
A_{\mu\nu} = -4 \nabla_\alpha\nabla_\beta C_{\mu}{}^{\alpha}{}_{\nu}{}^\beta
-2 C_\mu{}^\alpha{}_\nu{}^\beta \, R_{\alpha\beta}   \,.    \label{eq:A}
\end{equation}
The derivative terms lead to a potential problem of making the
Einstein equations
fourth-order equations and leading to unstable solutions. This effect
is analogous to the runaway solutions of the Lorentz-Dirac equation
for classical charged particles. Various solutions to this problem
have been suggested, including order-reduction approaches~\cite{PS93},
and criteria for the validity of the semiclassical theory~\cite{AMM03,AMM09}.

A well-known aspect of quantum stress tensor is the conformal anomaly.
At the classical level, the stress tensor of a conformally invariant
field has a vanishing trace. This no longer holds for the renormalized
stress tensor, where $\langle  T^\mu _\mu \rangle
\not=0$. Furthermore, the anomalous trace for a free field is a
state independent local geometric quantity which is quadratic in
the Riemann tensor. In the case of a conformally invariant field
in a conformally flat spacetime, the
unambiguous part of the anomalous trace arises from a geometrical
term in
$\langle  T_{\mu \nu} \rangle$ of the form $ C\,
\mathscr{B}_{\mu\nu}$,
where $C$ is a constant which depends upon the specific field, and
\begin{equation}
    \mathscr{B}_{\mu\nu}=- 2C_{\alpha\mu\beta\nu}R^{\alpha\beta} +
 \frac{1}{2}\,g_{\mu\nu}R_{\alpha\beta}R^{\alpha\beta}+
\frac{2}{3}\,R_{\mu\nu}R-R_{\mu}{}^{\alpha}R_{\nu\alpha}-
\frac{1}{4}\,g_{\mu\nu}R^{2}\,, \label{eq:B}
\end{equation}
where $C_{\alpha\mu\beta\nu}$ is the Weyl tensor. The term containing
the Weyl tensor vanishes in conformally flat spacetime, but is needed
to give the correct generalization to non-conformally flat spacetimes.
The tensor $ \mathscr{B}_{\mu\nu}$   was obtained by
Davies~{\it et al}~\cite{Davies77} and
by Bunch~\cite{Bunch}. The conformal anomaly is given by
\begin{equation}
\langle  T^\mu_\mu \rangle = C\, \mathscr{B}^\mu_\mu =
C\,\left( R_{\alpha\beta}R^{\alpha\beta} -\frac{1}{3} R^2\right)\,.
\end{equation}
More generally, there can be a term proportional to
 $C_{\mu\nu\alpha\beta} C^{\mu\nu\alpha\beta}$ in the anomalous
trace, but this term will vanish for weakly perturbed conformally
flat spacetime, such as we consider.

The semiclassical Einstein equations for gravity with a cosmological
constant $\Lambda$ coupled to a quantum field can be written as
\begin{equation}
R_{\mu\nu} - \Lambda g_{\mu\nu} = 8 \pi G
\left(\langle  T_{\mu \nu} \rangle- \frac{1}{2}g_{\mu\nu}
\langle  T^\rho_\rho \rangle \right)  \label{eq:Einstein} \,.
\end{equation}
In addition to the local,
geometric terms in $\langle  T_{\mu \nu} \rangle$, in general there
are non-local terms which are difficult to compute
explicitly. Fortunately, for the case of small perturbations around
a conformally flat spacetime, they have been found in
Refs.~\cite{S81,CV94,HW82}.  Here we will follow
the coordinate space formulation given by Horowitz and Wald~\cite{HW82},
which is based on earlier work by Horowitz~\cite{H80} and by Horowitz
and Wald~\cite{HW80}.

 To first order in the perturbation $h_{\mu\nu}$, Horowitz and Wald's
result can be written
as
\begin{equation}
\langle  T_{\mu \nu} \rangle = \beta\, H_{\mu\nu} +
C\,\mathscr{B}_{\mu\nu} + P_{\mu\nu} + Q_{\mu\nu}\,.  \label{eq:Tmunu}
\end{equation}
Here
\begin{equation}
P_{\mu\nu} = -16\pi \alpha\, a^{-2}\, \partial^\rho \partial^\sigma
[\ln(a)\, \tilde{C}_{\mu\rho\nu\sigma} ]\,,
\end{equation}
where $\tilde{C}_{\mu\rho\nu\sigma}$ is the Weyl tensor for perturbed
Minkowski spacetime with the perturbation
 $\tilde{h}_{\mu\nu} = a^{-2} h_{\mu\nu}$, the
partial derivatives are with respect to the Minkowski space
coordinates, and $\alpha$ is another constant which depends upon the
quantum field. The most complicated term in Eq.~(\ref{eq:Tmunu})
is the non-local part given by
\begin{equation}
Q_{\mu\nu} = \alpha\,  a^{-2}\, \int d^4 x'\, H_\lambda (x-x') \,
 \tilde{A}_{\mu\nu} \,,
\end{equation}
where
\begin{equation}
 \tilde{A}_{\mu\nu} = -4 \,  \partial^\rho \partial^\sigma
\, \tilde{C}_{\mu\rho\nu\sigma}\,,
\end{equation}
is the first order form of $A_{\mu\nu}$  for perturbed
Minkowski spacetime with the perturbation
$ \tilde{h}_{\mu\nu} = a^{-2} h_{\mu\nu}$.
The action of the distribution $H_\lambda(x-x')$ on a function $f$
can be expressed in terms of radial
null coordinates $u=t-r$ and $v=t+r$ and an angular integration as
\begin{equation}
\int d^4 x'\, H_\lambda (x-x') \, f(x') = \int_{-\infty}^0 du \int
d\Omega \left[ \frac{\partial f}{\partial u} \Biggl|_{v=0}\;
\ln(-u/\lambda) +\frac{1}{2} \frac{\partial f}{\partial v}
\Biggl|_{v=0} \right]\,.  \label{eq:H_lambda}
\end{equation}
This expression is an integral over the past lightcone of the point
$x$.

The result for $\langle  T_{\mu \nu} \rangle$, Eq.~(\ref{eq:Tmunu}),
contains two constants, $C$ and $\alpha$, whose values can be
determined explicitly, and are given in Table~\ref{table:C}
for several fields.
The remaining two constants, $\beta$ and $\lambda$, are undetermined.
A shift in either of these constants adds additional terms
proportional to  $H_{\mu\nu}$ and
$A_{\mu\nu} = a^{-2}  \tilde{A}_{\mu\nu}$, respectively.
We could have added a term of the form
$c_A \, A_{\mu\nu}$ to the right-hand side of
 Eq.~(\ref{eq:Tmunu}). The result would then be invariant
under changes in  $\lambda$ in the sense that a shift in
 $\lambda$ would alter $c_A$.

\begin{table}
\begin{tabular}{|c||c||c|} \hline\hline
Field & $C$ & $\alpha$  \\ \hline
Conformal scalar   & ${1}/({2880\, \pi^2})$ &  $1/(3840\, \pi^3)$   \\
Spin $\frac{1}{2}$ & ${11}/({5760\, \pi^2})$ & $1/(1280\, \pi^3)$ \\
Photon             & ${31}/({1440\, \pi^2})$  & $1/(320\, \pi^3)$ \\
\hline\hline
\end{tabular}
  \caption{The coefficients $C$ and $\alpha$ are listed for three
different massless
  fields where the spin $\frac{1}{2}$ field is the result for Weyl
fermions and becomes a factor of 2 larger for 4-component Dirac
fermions. This table is based on data from Refs.~\cite{H80,BD82}.}
\label{table:C}
\end{table}

\section{Effects of the Local Geometric Terms}
\label{sec:geom}

Here we treat the local, geometric tensors $H_{\mu\nu}$
and $ \mathscr{B}_{\mu\nu}$, and show that they produce no effects
on the tensor perturbations other than finite shifts of the
cosmological and Newton's constants. Write Eq.~(\ref{eq:Einstein})
as
\begin{equation}
R_{\mu\nu} - \Lambda_0 g_{\mu\nu} = 8 \pi\, G_0
\left(\langle  T_{\mu \nu} \rangle- \frac{1}{2}g_{\mu\nu}
\langle  T^\rho_\rho \rangle \right)  \, , \label{eq:Einstein2}
\end{equation}
where $\Lambda_0$ and $G_0$ are the cosmological and Newton's
constants after all infinite renormalizations have occurred, but before
these finite shifts. Here we take
\begin{equation}
\langle  T_{\mu \nu} \rangle = \beta\, H_{\mu\nu} +
C\,\mathscr{B}_{\mu\nu}\,.  \label{eq:Tmunu_local}
\end{equation}

To zeroth order, that is, on the de~Sitter background, we have
\begin{equation}
{}^{(0)}\mathscr{B}_{\mu\nu}
=-\frac{1}{3}\,\gamma_{\mu\nu}\Lambda^2\,, \quad
{}^{(0)}\mathscr{B}=-\frac{4}{3}\,\Lambda^2\,, \quad {\rm and }\;
{}^{(0)}H_{\mu\nu} = 0\,.  \label{eq:zeroth}
\end{equation}
If we insert these relations in  Eq.~(\ref{eq:Einstein}), we find
\begin{equation}
{}^{(0)}R_{\mu\nu} - \Lambda \gamma_{\mu\nu} = 0\,  , \label{eq:Ricci0}
\end{equation}
where shifted cosmological constant, $\Lambda$, is related to  $\Lambda_0$
by
\begin{equation}
\Lambda = \Lambda_0 + \frac{8 \pi}{3} G_0\, C\, \Lambda^2
\,  , \label{eq:Lambda}
\end{equation}
In general, $ \mathscr{B}_{\mu\nu}$ is not of the form of a
cosmological constant term, but in de~Sitter space, it produces an
effective shift in  $\Lambda$. Here we have written the second
term on the right-hand side of
Eq.~(\ref{eq:Lambda}) in terms of the shifted cosmological constant,
$\Lambda$, but to the order we are working, we could have equally well
used $\Lambda_0$.

Next we need to find the explicit forms for the various tensors in
Eq.~(\ref{eq:Einstein2}) to first order in $h_{\mu\nu}$ in the transverse,
trace-free gauge, Eq~(\ref{eq:TT}). The Ricci tensor
has the first order form
\begin{equation}
	{}^{(1)}R_{\mu\nu}=-\frac{1}{2}  h_{\mu\nu;\alpha}^{~~~~~\alpha} +
\frac{4}{3}\,\Lambda \,h_{\mu\nu}\, .  \label{eq:R1}
\end{equation}
Thus, if $\langle  T_{\mu \nu} \rangle=0$, Eq.~(\ref{eq:Einstein})
becomes $h_{\mu\nu;\alpha}^{~~~~~\alpha} - \frac{2}{3}\,\Lambda \,h_{\mu\nu}
=0$, which is equivalent to Eq.~(\ref{eq:scalar}).
Note that in general, the transverse, trace-free gauge cannot be
imposed in the presence of sources. However, here all the terms in
the first order Einstein equations are traceless, so this gauge may
be used consistently. (Strictly,  it is $^{(1)}R^\mu_\nu$ which is
a gauge invariant quantity, whereas  $^{(1)}R^{\mu\nu}$ and
$^{(1)}R_{\mu\nu}$ are not necessarily gauge invariant~\cite{SW}.)
The first order form of $H_{\mu\nu}$ is
\begin{equation}
{}^{(1)}H_{\mu\nu} = 4 \Lambda\, \left( h_{\mu\nu;\alpha}^{~~~~~\alpha} -
 \frac{2}{3}\,\Lambda \,h_{\mu\nu} \right)\, , \label{eq:H1}
\end{equation}
and that of  $ \mathscr{B}_{\mu\nu}$ is
\begin{equation}
{}^{(1)}\mathscr{B}_{\mu\nu}=-\frac{1}{3}\Lambda \left(
  h_{\mu\nu;\alpha}^{~~~~~\alpha}  +\frac{1}{3}\Lambda \,h_{\mu\nu}\right)\,.
\end{equation}
The net contribution of $ \mathscr{B}_{\mu\nu}$ to the right hand side of
Eq.~(\ref{eq:Einstein2}) is proportional to
\begin{equation}
{}^{(1)}\mathscr{B}_{\mu\nu}- \frac{1}{2}\,h_{\mu\nu}  {}^{(0)}\mathscr{B}
= -\frac{1}{3}\Lambda \left(
  h_{\mu\nu;\alpha}^{~~~~~\alpha}  -\frac{5}{3}\Lambda
  \,h_{\mu\nu}\right)\,. \label{eq:B1}
\end{equation}
If we use Eqs.~(\ref{eq:Lambda}), (\ref{eq:R1}), (\ref{eq:H1}), and
(\ref{eq:B1}),
then we may write Eq.~(\ref{eq:Einstein2}) as
\begin{equation}
\left(1 + 64 \pi G_0 \, \beta\,\Lambda  -  \frac{16 \pi}{3} G_0\, C\, \Lambda\right)
({}^{(1)}R_{\mu\nu} - \Lambda h_{\mu\nu} ) = 0\,  .
\end{equation}
This implies that once we introduce additional terms in the stress
tensor, the Einstein equation becomes Eq.~(\ref{eq:Einstein}), with
the shifted Newton's constant given by
\begin{equation}
G = \ell_p^2 =
G_0\;\left(1 + 64 \pi G_0 \, \beta\,\Lambda  -  \frac{16 \pi}{3} G_0\,
C\, \Lambda \right)^{-1}
\,  , \label{eq:G}
\end{equation}
where $\ell_p^2$ is the Planck length.

Now we may consider only the effects of the $P_{\mu\nu}$ and $Q_{\mu\nu}$ terms
on the tensor perturbations, which satisfy the equation
\begin{equation}
\Box_s h_i ^j = -16 \pi \ell_p^2\, (P_i^j + Q_i^j)\,, \label{eq:tensor}
\end{equation}
in the transverse, trace-free gauge.

\section{Spatially Homogeneous Solutions}
\label{sec:stab}

In this section, we study the stability of the tensor perturbations of de~Sitter spacetime
in the presence of the quantum stress tensor of the conformal field. For this purpose,
it is sufficient to examine spatially homogeneous solutions of Eq.~(\ref{eq:tensor}), as
these will be the most rapidly growing modes if there is an
instability. Note that the tensor modes which we are considering are
associated with anisotropic perturbations, even when they are
 spatially homogeneous. This follows from the fact that they have
non-vanishing Weyl tensor. Thus, the results of this section are
distinct from, but complementary to, recent results by
 P{\'e}rez-Nadal {\it et al}~\cite{PNRV08}, who demonstrate stability
of de Sitter spacetime under isotropic perturbations at the one-loop
level in semiclassical gravity.

In order to find the tensors $P_{\mu\nu}$ and $Q_{\mu\nu}$, we first
need $\tilde{C}_{\mu\rho\nu\sigma}$. We here ignore spatial derivatives,
and restrict our attention to spatial components, which are the only
nontrivial ones in our gauge. Then we
need $\tilde{A}_{ij} = -4\, \tilde{C}_{i\eta j \eta,\eta \eta}$.
The relevant components of the Riemann and Ricci tensors associated
with the Minkowski perturbation $\tilde{h}_{ij}$ are $\tilde{R}_{i\eta j \eta} =
 -\frac{1}{2} \tilde{h}_{ij,\eta \eta}$ and  $\tilde{R}_{i j} =
 \frac{1}{2} \tilde{h}_{ij,\eta \eta}$. Note that although  ${h}_{ij}$
is a gravity wave on de~Sitter spacetime,  $\tilde{h}_{ij}$ is not a
source-free solution near flat spacetime. From these results, we
obtain $ \tilde{C}_{i\eta j \eta,\eta \eta} = \tilde{R}_{i\eta j \eta}
+  \frac{1}{2} \tilde{R}_{i j} = -\frac{1}{4} \tilde{h}_{ij,\eta \eta}$
and hence
\begin{equation}
\tilde{A}_{ij} =\partial^4_\eta\, \tilde{h}_{ij}\,.
\end{equation}
We may express the local tensor $P_{ij}$ as
\begin{equation}
P_{ij} = 4\pi \alpha\, a^{-2} \left[\ln(a)\, ( a^{-2}
  \,{h}_{ij})_{,\eta\eta} \right]_{,\eta\eta} \,. \label{eq:P}
\end{equation}
The non-local term involves the distribution $H_\lambda$, and
an integral over the past lightcone of the point $x$ at which the
stress tensor is evaluated. Take $r=0$ at this point, in which case
we may write $u=\eta'-\eta -r'$ and  $v=\eta'-\eta +r'$. The
function on which the distribution acts depends only upon $\eta'$, so
$f = f(\eta') = f[\frac{1}{2} (u+v) +\eta]$. As a result,
$(\partial f/\partial u)_{v=0} = (\partial f/\partial v)_{v=0}
= \frac{1}{2} f'(\eta')$, and we may write Eq.~(\ref{eq:H_lambda}) as
\begin{equation}
\int d^4 x'\, H_\lambda (x-x') \, f(\eta') = 4\pi\int_{-\infty}^0 d\eta'
 \left\{ f'(\eta') \,\ln \left[ \frac{2(\eta-\eta')}{\lambda} \right]
+ \frac{1}{2} f'(\eta') \right\}\,.
\end{equation}
The last term in the integrand may be absorbed in a redefinition of
$\lambda$, and hence will be dropped. Thus we obtain
\begin{equation}
Q_{ij} = 4\pi \alpha\, a^{-2} \, \int_{-\infty}^0 d\eta' \,
 \partial^5_{\eta'} \tilde{h}_{ij}(\eta') \,
\ln \left[ \frac{2(\eta-\eta')}{\lambda} \right]\,.  \label{eq:Q}
\end{equation}

We wish to look for a growing, spatially homogeneous solution of
Eq.~(\ref{eq:Einstein}). In particular, let
\begin{equation}
\tilde{h}_{ij} =  a^{-2}\, {h}_{ij} = h_i^j = e_i^j\, (-\eta)^{-b} \,,
\label{eq:soln}
\end{equation}
where $e_i^j$ is a constant tensor and $b$ is a constant. A solution
for which $b>0$ will grow as a power of conformal time as
$\eta \rightarrow 0$, or exponentially in comoving time.

If we insert Eq.~(\ref{eq:soln}) into Eq.~(\ref{eq:P}), the result is
\begin{equation}
P_i^j = 4\pi \alpha\, e_i^j\, H^4\, (-\eta)^{-b}\, b(1+b) [2b +5
-(2+b)(3+b)\, \ln(-H \eta)]\,. \label{eq:P2}
\end{equation}
Similarly, Eq.~(\ref{eq:Q}) yields
\begin{equation}
Q_i^j = 4\pi \alpha\, e_i^j\, H^4\, (-\eta)^{-b}\, b(1+b)(2+b)(3+b)
[ \ln(-2 \eta/\lambda) -\psi(b+4) -\gamma ] \, , \label{eq:Q2}
\end{equation}
where $\gamma$ is Euler's constant and $\psi$ is the digamma function.
The scalar wave operator in de~Sitter spacetime here has the form
\begin{equation}
\Box_s h_i ^j = -H^2 \eta^4 \frac{d}{d \eta} \left(\eta^{-2} \, \frac{d}{d \eta} \right)\,h_i ^j \,.
\end{equation}
Equation~(\ref{eq:tensor}) may now be written as
\begin{equation}
b(3+b) = -\xi\, (2+b)(3+b)\,\left\{ b(1+b)\;[\psi(b) + \gamma +\ln(H \lambda/2)] +
1+2b \right\}\,, \label{eq:b}
\end{equation}
where $\xi = 64 \pi^2 \, \ell_p^2\,H^2 \alpha$, and we have used the identity
 $\psi(x+1) =\psi(x) + 1/x$. Thus the homogeneous solutions in the absence
of the quantum stress tensor ($\xi=0$) are $b=0$ and $b=-3$, which are both stable.
The only possibility for an unstable solution which is within the domain of validity of
the semiclassical theory is one with a small positive value of $b$. If we expand
Eq.~(\ref{eq:b}) for $|b| \ll 1$, we find
\begin{equation}
b(3+b) \approx -\xi\, \left\{ 6[1 +\ln(H \lambda/2)]\,b +
[5+\pi^2 +11 \ln(H \lambda/2)] \, b^2 +O(b^3) \right\} \,.
\end{equation}
Thus $b=0$ is still a solution, and the second solution will
be $b \approx -3$ so long as $\xi \ll 1$ and  $\xi \, |\ln(H
\lambda/2)| \ll 1$.
These latter conditions can be considered to be
criteria for the validity of the semiclassical theory.
 Hence we conclude that de~Sitter spacetime is stable in the
semiclasical theory against tensor
perturbations. Here we should comment on the explicit appearance of
the parameter $\lambda$ in Eq.~(\ref{eq:b}). Although the theory is
invariant under changes in  $\lambda$ so long as there is a term
proportional to  $A_{\mu \nu}$ in $\langle  T_{\mu \nu} \rangle$, we
have set the coefficient of this term to zero, which is analogous to
a gauge choice. In any case, our conclusion does not depend upon the
value of  $\lambda$ in Eq.~(\ref{eq:b}), so long as $\xi \, |\ln(H
\lambda/2)| \ll 1$.  If this condition is not fulfilled, any
resulting instabilities can be
viewed as a breakdown of the semiclassical theory.

\section{Effects on Gravity Waves}
\label{sec:correct}

In this section, we will study the effect of the quantum stress tensor on gravity waves in
de~Sitter spacetime. The plane wave solutions of Eq.~(\ref{eq:scalar}) are of the form
\begin{equation}
h_\mu ^\nu = c_0\, e_\mu ^\nu \,(1+i k \eta)\,
{\rm e}^{i(\mathbf{k} \cdot \mathbf{x} - k \eta)}\,,
  \label{eq:grav_wave}
\end{equation}
where $c_0$ is a constant and $ e_\mu ^\nu$ is the polarization tensor. We need
to compute the quantum stress tensor in perturbed de~Sitter spacetime, with this
plane wave perturbation. The first step in finding the tensors  $P_{\mu\nu}$ and
$Q_{\mu\nu}$ is constructing $\tilde{C}_{\mu\rho\nu\sigma}$, the Weyl tensor associated
with the conformally transformed perturbation of flat spacetime, $\tilde{h}_{\mu\nu}$.
Note that mixed components of $\tilde{h}_\mu^\nu$ coincide with those of the original
perturbation of de~Sitter spacetime, ${h}_\mu^\nu$. However, $\tilde{h}_\mu^\nu$ is
not a vacuum solution of perturbed flat space, and has a non-zero Ricci tensor
\begin{equation}
\tilde{R}_\mu^\nu = -\frac{1}{2} \tilde{\Box} \tilde{h}_\mu^\nu\,,
\end{equation}
where $ \tilde{\Box} $ is the flat space wave operator. Similarly, we find the associated
Riemann tensor to satisfy
\begin{equation}
\partial^\rho \partial^\sigma\, \tilde{R}_{\mu\rho\nu\sigma} =
 -\frac{1}{2} \tilde{\Box}  \tilde{\Box} \tilde{h}_{\mu\nu} \,.
\end{equation}
Hence the tensor $\tilde{A}_{\mu\nu}$ and the Weyl tensor satisfy
\begin{equation}
\tilde{A}_{\mu\nu}  = -4  \partial^\rho \partial^\sigma\, \tilde{C}_{\mu\rho\nu\sigma} =
 \tilde{\Box}  \tilde{\Box} \tilde{h}_{\mu\nu} \,.
\end{equation}
However, when we use the perturbation given by Eq.~(\ref{eq:grav_wave}), we find
that $\tilde{A}_{\mu\nu}  = 0$, so the non-local term vanishes:
\begin{equation}
{Q}_{\mu\nu}  = 0\,.
\end{equation}
The tensor $P_{\mu\nu}$ is non-zero and is given by
\begin{equation}
P_\mu^\nu  =  8 \pi i \alpha H^2\, e_\mu ^\nu \, c_0 k^3\, \eta\,
{\rm e}^{i(\mathbf{k} \cdot \mathbf{x} - k \eta)} \, .
\end{equation}

In the presence of the quantum stress tensor, the modified gravity wave may
be expressed as $h_\mu^\nu + {h'}_\mu^\nu$, where
\begin{equation}
{h'}_\mu^\nu(x) = 16 \pi \ell_p^2 \int d^4 x' \, \sqrt{-g(x')}\, G_R(x,x')\; P_\mu^\nu \, ,
\end{equation}
where $G_R(x,x')$ is the scalar retarded Green's function in de~Sitter space.
This function vanishes for $\eta < \eta'$ and satisfies
\begin{equation}
\Box_s\, G_R(x,x') = - \frac{\delta(x-x')}{\sqrt{-g(x')}} \,.
\end{equation}
It is convenient to take a spatial Fourier transform and write
\begin{equation}
 G_R(x,x') = \frac{1}{a^2(\eta')\, (2\pi)^3}\; \int d^3 k \,
 {\rm e}^{i \mathbf{k} \cdot (\mathbf{x} - \mathbf{x'})} \; G(\eta,\eta';k) \, ,
\end{equation}
where $G(\eta,\eta';k)$ satisfies
\begin{equation}
\frac{d^2 G}{d \eta^2} - \frac{2}{\eta} \, \frac{d G}{d \eta} + k^2  G
= \delta(\eta-\eta') \,.    \label{eq:G_eq}
\end{equation}
The explicit form for $G(\eta,\eta';k)$ is given in Eq.~(72) of Ref.~\cite{WKF07}, and
may be expressed as
\begin{equation}
G(\eta,\eta';k) = \frac{1}{k^3\, (\eta')^2} \; \left\{ (1+k^2 \eta \eta')\, \sin[k (\eta -\eta')]
- k (\eta -\eta')\, \cos[k (\eta -\eta')] \right\} \,,
\end{equation}
for $\eta > \eta'$.

Suppose that the effect of the quantum stress tensor is switched on at
time $\eta = \eta_0$, which we could take to be the beginning of inflation.
Now we may write the solution for ${h'}_\mu ^\nu(x)$, which vanishes for $\eta < \eta_0$,
as
\begin{eqnarray}
{h'}_\mu^\nu(x) &=& 128 \pi^2\, i  e_\mu ^\nu \,c_0 \alpha \, H^2 \,\ell_p^2
   {\rm e}^{i \mathbf{k} \cdot \mathbf{x}} \nonumber \\
&\times& \int_{\eta_0}^\eta d\eta'  \left\{ (1+k^2 \eta \eta')\, \sin[k (\eta -\eta')]
- k (\eta -\eta')\, \cos[k (\eta -\eta')] \right\} \, \frac{ {\rm  e}^{-ik \eta'}}{\eta'} .
\label{eq:mod_int}
\end{eqnarray}
In the limit that $\eta_0 \rightarrow - \infty$, that is, $|\eta_0|
\gg |\eta|$, the dominant contribution
to the integral will come from terms in the integrand which are independent of $\eta'$.
This leads to a result proportional to  $|\eta_0|$,
\begin{equation}
{h'}_\mu^\nu(x) \sim - 64 \pi^2\, i \, e_\mu ^\nu \,c_0 \alpha\, H^2 \,k \ell_p^2\, |\eta_0|
 \,(1+i k \eta)\,{\rm e}^{i(\mathbf{k} \cdot \mathbf{x} - k \eta)}\,,
  \label{eq:mod_wave}
\end{equation}
which has the same functional form as does $h_\mu^\nu$, but is out of
phase by $-\pi/2$ due to the factor of $-i$.

The most striking feature of the result Eq.~(\ref {eq:mod_wave}) is that the correction term
due to the quantum stress tensor is proportional to $|\eta_0|$, and hence is larger the
earlier the coupling between the quantum stress tensor and the metric perturbation is
switched on. This bears some similarities to the results found in
Refs.~\cite{WKF07,FMNWW10}, where the effects of conformal stress tensor fluctuations
in inflation were found to depend upon powers of the scale factor change during inflation.
However, here we are concerned with an effect of the stress tensor expectation value,
and not with fluctuations around this value. In all cases, the dependence upon $|\eta_0|$
might appear to violate a theorem first due to
Weinberg~\cite{Weinberg}. (See also Ref.~\cite{Chai07}.) This result
states that quantum loop effects should grow no faster than logarithmically with the scale factor
during inflation. However, it was argued in Ref.~\cite{FMNWW10} that there is no real
violation of this theorem, because the quantum effects are not so much growing as always
large, and are due to very high frequency modes at $\eta = \eta_0$.

The same interpretation applies to our present result, Eq.~(\ref{eq:mod_wave}). We can
write the ratio of the magnitude of the correction to that of the original wave as
\begin{equation}
\Gamma = \left | \frac{{h'}_\mu^\nu}{{h}_\mu^\nu}\right |  =
64 \pi^2 \alpha H^2 k \ell_p^2 |\eta_0| =  64 \pi^2 \alpha H\, k_P\, \ell_p^2 \,.
\label{eq:Gamma}
\end{equation}
Here $k_P = k/a(\eta_0) = k H  |\eta_0| $ is the physical wavenumber of the mode
as measured by a comoving observer at time $\eta = \eta_0$. If we require that the
curvature of the de~Sitter spacetime be well below the Planck scale, then we have
\begin{equation}
H\, \ell_p \ll 1 \,.
\end{equation}
Similarly, if the mode in question is always below the Planck scale while it interacts
with the quantum stress tensor, then
\begin{equation}
k_P\, \ell_p \ll 1 \,.
\end{equation}
These two conditions together imply that $|{{h'}_\mu^\nu}/{{h}_\mu^\nu}| < 1$, and
hence the quantum correction to the gravity wave is smaller than the original wave.
Another possibility it that we should take transplanckian mode
seriously, and allow their contributions. This issue was discussed in
more detail in Ref.~\cite{FMNWW10}, where numerous references to
earlier papers may be found.

Note that the dependence upon $\eta_0$ does not arise from sudden
switching at $\eta = \eta_0$, but only from the duration of inflation
in conformal time. A more precise form for ${{h'}_\mu^\nu}$ is
obtained by replacing $\eta_0$ by $\eta_0 -\eta$ in
Eq.~(\ref{eq:mod_wave}). Thus the modified wave is no longer exactly 
a solution of the Lifshitz equation, Eq.~(\ref{eq:scalar}). It is
no longer constant when the mode has a proper wavelength larger than
the horizon size, $k |\eta| < 1$. This is in contrast to the
unperturbed mode, Eq.~(\ref{eq:grav_wave}), whose magnitude is
approximately constant when it is outside the horizon.

\section{Tensor Perturbations in Inflationary Cosmology}
\label{sec:tensor-power}

One of the successes of inflationary cosmology is the prediction of a
Gaussian and nearly scale invariant spectrum of primordial density
fluctuations~\cite{MC81,GP82,Hawking82,Starobinsky82,BST83},
which seems to be confirmed by measurements on the cosmic microwave
background (CMB)~\cite{WMAP}.  Another prediction is a similar spectrum
of tensor perturbations, which might be found in polarization
measurements of the CMB, but at the present these perturbations have not
been detected.

The tensor perturbations from inflation are less model dependent than
are the density perturbations, The former arise from vacuum modes
of the quantized graviton field in de Sitter spacetime which evolve
according to the Lifshitz equation, Eq.~(\ref{eq:scalar}), until
the last scattering surface. At this time, they leave an imprint
on the CMB in the form of a power spectrum of tensor perturbations
given by (see, for example, Ref.~~\cite{Mukhanov}.)
\begin{equation}
\delta^2_h \approx \frac{8}{\pi}\, \ell_p^2\, H^2 \,.
\label{eq:flat_spec}
\end{equation}
This is an approximately flat spectrum. If $H$ slowly decreases as inflation progresses, then the spectrum is slightly enhanced for longer wavelengths. The numerical coefficient is fixed by the normalization of vacuum
graviton modes, which leads to $c_0 = \ell_p \sqrt{16 \pi/k}$
in Eq.~(\ref{eq:grav_wave}).

The effect of the conformal stress tensor is to modify the amplitude
of these modes by a factor of $1 - i \Gamma$, where $\Gamma$ is
given by Eq.~(\ref{eq:Gamma}). This in turn multiplies the power
spectrum by $|1-i \Gamma|^2 = 1 +\Gamma^2$. In order to estimate
this enhancement factor, we need to make some assumptions about a model
of inflation. Let $E_R$ be the reheating energy and assume that most
of the vacuum energy which drives inflation is converted into radiation at
reheating. Then Einstein's equations yield
\begin{equation}
H^2 = \frac{8 \pi}{3} \, \ell_p^2\, E_R^4 \,.
\end{equation}
For this discussion, we assume that $H$ is approximately constant
throughout the inflationary era.
There is expansion by a factor of about $E_R/({1\,\rm eV})$ between the
end of inflation and last scattering and a further expansion by a
factor of $10^3$ to the present. Let us choose the scale factor to be
unity at the end of inflation, so its present value will be
\begin{equation}
a_{\rm now} = 10^3 \frac{E_R}{1\,\rm eV} \,.
\end{equation}
Consider a scale which presently has a proper length of $\ell_0$, and
hence a physical wavenumber of $k_P = 2 \pi/\ell_0$. At the end of
inflation, its physical and comoving wavenumber coincide and are
given by
\begin{equation}
k = \frac{2 \pi\, a_{\rm now}}{\ell_0} \,.
\end{equation}
Recall that $k$ is constant, so this form holds throughout the
cosmological expansion.

Let
\begin{equation}
S = H \,|\eta_0|\,,
\end{equation}
which is the factor by which the universe expands from the initial
conformal time $\eta = \eta_0$ to the end of inflation. We may combine
the above relations to write
\begin{equation}
\Gamma^2 = \frac{8 \pi}{3} \,(128 \pi^3 \alpha)^2 \ell_p^6\, E_R^4\,
S^2\, \frac{a_{\rm now}^2}{\ell_0^2} \,.
\end{equation}
If we use the value of $\alpha = 1/(320 \pi^3)$ corresponding to the
electromagnetic field, then we may write
\begin{equation}
\Gamma^2 = 1.34 \times 10^{-78} \;
\left(\frac{10^{25}\,{\rm  cm}}{\ell_0}\right)^2\,
\left(\frac{E_R}{10^{15}\,{\rm GeV}} \right)^6 \; S^2 \,.
\label{eq:Gamma2}
\end{equation}
Recall that the present horizon size is of order $10^{28}\,{\rm  cm}$,
so $\ell_0\,\approx 10^{25} {\rm  cm}$ corresponds to angular scales of
the order of $1^\circ$ today.

If one has only the minimal inflation needed to solve the horizon and
flatness problems, so $S \approx 10^{23}$, then the effects of the
one-loop correction on the tensor perturbation spectrum is negligible.
However, larger values of $S$ have the potential to produce significant
corrections. For example, $E_R \approx 10^{15} {\rm GeV}$ and
$S \approx 10^{39}$ would lead to an effect of order unity at
$1^\circ$ scales. One should expect the one-loop approximation to
begin to break down, but this can serve as an order of magnitude
estimate. In contrast to the nearly flat spectrum,
Eq.~(\ref{eq:flat_spec}), due to free graviton fluctuations, the
one-loop effect is highly tilted toward the blue end of the spectrum.

It is of interest to compare the magnitude of this effect on the
tensor perturbations with the stress tensor fluctuation effect on
density perturbations which was treated in Refs.~\cite{WKF07,FMNWW10}.
The latter effect becomes significant if  $E_R \approx 10^{15} {\rm GeV}$ and
$S \approx 10^{33}$ (See Eq.~(108) in Ref.~~\cite{FMNWW10}.), and
is hence somewhat larger than the effect treated in the present paper.

\section{Summary}
\label{sec:final}

We have constructed the semiclassical Einstein equation with a
conformal matter field on a weakly
perturbed de~Sitter background,  using the coordinate space formulation
of  Horowitz and Wald~\cite{H80,HW80,HW82}, and  examined gravity wave
solutions of this equation. We found no growing, spatially homogeneous
(but anisotropic) solutions in a spatially, flat universe, which
implies that de~Sitter spacetime is stable to tensor perturbations
at the one-loop level in the presence of conformal matter.

We further examined the effects of the one-loop correction on the
propagation of finite wavelength gravity waves, and found a correction
term which depends upon the interval over which the interaction with
the quantum matter field is switched on, or equivalently, the duration
of inflation. So long as the curvature of de~Sitter spacetime and the
initial proper frequency of the mode are below the Planck scale, the
fractional correction is small. The effect take the form of both a
phase shift and an amplitude change. If one is concerned only
with the form of the gravity wave modes at late times, this effect can
be absorbed in a complex amplitude shift. However,
 gravity wave modes are no longer exactly
solutions of the Lifshitz equation, Eq.~(\ref{eq:scalar}).

The effect is potentially observable with a sufficient amount
of inflation through an increase in the amplitude of the spectrum
of tensor perturbations of the cosmic microwave background. This
possibility does require one to take seriously the contribution
of modes which were transplanckian at the beginning of inflation.

\begin{acknowledgments}
We would like to thank A. Higuchi, E. Mottola, A. Roura, and E. Verdaguer for
valuable discussions.
 This work is partially supported by the National Center for Theoretical
 Sciences, Taiwan, by Grants NSC 97-2112-M-001-005-MY3 and
NSC 97-2112-M-259-007-MY3, and by the U.S. National Science Foundation under
Grant PHY-0855360.  LHF would like to thank Academia Sinica and
National Dong Hwa University for hospitality while this work was conducted.
\end{acknowledgments}

\end{document}